\documentclass[twocolumn,showpacs,amssymb]{revtex4}

\usepackage{epsfig}
\usepackage{url}
\usepackage{longtable}
\usepackage{rotating}
\usepackage{natbib,times}
\usepackage{threeparttable}
\usepackage{amsmath}
\usepackage{graphicx}
\usepackage{subfigure}

\newcommand\jcap{{J. Cosmology Astropart. Phys.}}

\makeatletter
\let\jnl@style=\rm
\def\ref@jnl#1{{\jnl@style#1}}

\def\aj{\ref@jnl{AJ}}                   
\def\araa{\ref@jnl{ARA\&A}}             
\def\apj{\ref@jnl{ApJ}}                 
\def\apjl{\ref@jnl{ApJ}}                
\def\apjs{\ref@jnl{ApJS}}               
\def\ao{\ref@jnl{Appl.~Opt.}}           
\def\apss{\ref@jnl{Ap\&SS}}             
\def\aap{\ref@jnl{A\&A}}                
\def\aapr{\ref@jnl{A\&A~Rev.}}          
\def\aaps{\ref@jnl{A\&AS}}              
\def\azh{\ref@jnl{AZh}}                 
\def\baas{\ref@jnl{BAAS}}               
\def\jrasc{\ref@jnl{JRASC}}             
\def\memras{\ref@jnl{MmRAS}}            
\def\mnras{\ref@jnl{MNRAS}}             
\def\pra{\ref@jnl{Phys.~Rev.~A}}        
\def\prb{\ref@jnl{Phys.~Rev.~B}}        
\def\prc{\ref@jnl{Phys.~Rev.~C}}        
\def\prd{\ref@jnl{Phys.~Rev.~D}}        
\def\pre{\ref@jnl{Phys.~Rev.~E}}        
\def\prl{\ref@jnl{Phys.~Rev.~Lett.}}    
\def\pasp{\ref@jnl{PASP}}               
\def\pasj{\ref@jnl{PASJ}}               
\def\qjras{\ref@jnl{QJRAS}}             
\def\skytel{\ref@jnl{S\&T}}             
\def\solphys{\ref@jnl{Sol.~Phys.}}      
\def\sovast{\ref@jnl{Soviet~Ast.}}      
\def\ssr{\ref@jnl{Space~Sci.~Rev.}}     
\def\zap{\ref@jnl{ZAp}}                 
\def\nat{\ref@jnl{Nature}}              
\def\iaucirc{\ref@jnl{IAU~Circ.}}       
\def\aplett{\ref@jnl{Astrophys.~Lett.}} 
\def\apspr{\ref@jnl{Astrophys.~Space~Phys.~Res.}}
\def\bain{\ref@jnl{Bull.~Astron.~Inst.~Netherlands}} 
\def\fcp{\ref@jnl{Fund.~Cosmic~Phys.}}  
\def\gca{\ref@jnl{Geochim.~Cosmochim.~Acta}}   
\def\grl{\ref@jnl{Geophys.~Res.~Lett.}} 
\def\jcp{\ref@jnl{J.~Chem.~Phys.}}      
\def\jgr{\ref@jnl{J.~Geophys.~Res.}}    
\def\jqsrt{\ref@jnl{J.~Quant.~Spec.~Radiat.~Transf.}}
\def\memsai{\ref@jnl{Mem.~Soc.~Astron.~Italiana}}
\def\nphysa{\ref@jnl{Nucl.~Phys.~A}}   
\def\physrep{\ref@jnl{Phys.~Rep.}}   
\def\physscr{\ref@jnl{Phys.~Scr}}   
\def\planss{\ref@jnl{Planet.~Space~Sci.}}   
\def\procspie{\ref@jnl{Proc.~SPIE}}   

\makeatother

\begin{document}

\title{Detection of gamma-ray  emission from the Coma cluster with {\it Fermi} Large Area Telescope and tentative evidence for an extended spatial structure}
\author{Shao-Qiang Xi$^{1,2}$ , Xiang-Yu Wang$^{1,2,*}$ , Yun-Feng Liang$^3$, Fang-Kun Peng$^{1,4}$, Rui-Zi Yang$^5$, Ruo-Yu Liu$^5$}
\affiliation{$^1$ School of Astronomy and Space Science, Nanjing
University, Nanjing 210093, China\\
$^2$Key laboratory of Modern Astronomy and Astrophysics, Nanjing
University, Ministry of Education, Nanjing 210093, China\\
$^3$ Purple Mountain Observatory, Chinese Academy of Sciences, Nanjing 210008, China\\
$^4$ School of Physics and Electronic Science, Guizhou normal University, Guiyang 550001, China \\
$^5$ Max-Planck-Institut f\"ur Kernphysik, Saupfercheckweg 1, 69117 Heidelberg, Germany\\
*Email: xywang@nju.edu.cn}

\begin{abstract}
Many galaxy clusters  have giant halos of non-thermal radio emission, indicating the presence of relativistic electrons in the clusters. Relativistic protons may also be accelerated by merger and/or accretion shocks in galaxy clusters. These cosmic-ray (CR) electrons and/or protons are expected to produce gamma-rays through inverse-Compton scatterings or inelastic $pp$ collisions respectively. Despite of intense efforts in searching for high-energy gamma-ray emission from galaxy clusters, conclusive evidence is still missing so far.  Here we report the discovery of $\ge 200$ MeV gamma-ray emission from the Coma cluster direction with an unbinned likelihood analysis of the 9 years of {\it Fermi}-LAT Pass 8 data.  The gamma-ray emission shows a spatial morphology roughly coincident with the giant radio halo, with an apparent excess at the southwest  of the cluster. Using the test statistic analysis, we further find tentative evidence that the gamma-ray emission at the Coma center is spatially extended. The extended component has an integral energy flux of  $\sim 2\times 10^{-12}{\rm \ erg\ cm^{-2}\ s^{-1}}$ in the energy range of 0.2 - 300 GeV and the spectrum is soft with a photon index of $\simeq-2.7$. Interpreting the gamma-ray emission as arising from CR proton interaction, we find that the volume-averaged value of the CR to thermal pressure ratio in the Coma cluster is about $\sim 2\%$. Our results show that  galaxy clusters are  likely a new type of GeV gamma-ray sources, and they are probably also giant reservoirs of CR
protons.
\end{abstract}
\maketitle

\section{ Introduction}
Galaxy clusters, the largest gravitationally bound structures in the Universe, are thought to form through mergers and accretion of smaller structures.  The merger-driven shocks and turbulence may accelerate  particles to relativistic energies (e.g. see \cite{2014IJMPD..2330007B} for a recent review). The presence of relativistic electrons in intra-cluster medium (ICM) as well as magnetic fields has been demonstrated by the detections of  Mpc-scale non-thermal synchrotron radio halos in many clusters \cite{2012A&ARv..20...54F}. CR proton acceleration has also been predicted in galaxy clusters \cite{1980ApJ...239L..93D,1999APh....12..169B,2009IJMPD..18.1609M,2004A&A...413...17P,2010ApJ...722..737K}, although at different levels for different acceleration scenarios.
Gamma-ray emission can be produced by the  neutral pion decay in the hadronic scenario, or by  inverse-Compton (IC) scatterings of the cosmic microwave background (CMB) photons on ultra-relativistic electrons \cite{2000Natur.405..156L,2001ApJ...562..233M,2007IJMPA..22..681B,2009JCAP...08..002K,2010MNRAS.409..449P}. In the re-acceleration model for radio halos (e.g., \cite{2001MNRAS.320..365B,2001ApJ...557..560P,2011MNRAS.410..127B, 2017MNRAS.472.1506B,2017MNRAS.465.4800P}), protons can also be accelerated  and consequently produce also gamma-rays, although at a level that is
smaller than that generally expected from the  pure hadronic models.  Thus, galaxy clusters have long been expected to be gamma-ray sources and the flux of  gamma-ray emission would provide  crucial constraints on the origin of the radio halos. Observationally, gamma-ray emission from clusters of galaxies were searched for a long time, including both individual cluster analysis and stacking procedures, but all these attempts resulted in non-detection or insignificant detection so far (e.g., \cite{2003ApJ...588..155R, 2010ApJ...717L..71A,2010JCAP...05..025A,2012ApJ...757..123A,2012arXiv1210.1574K,2012AAS...21920701Z, 2012MNRAS.427.1651H, 2012JCAP...07..017A, 2013A&A...560A..64H, 2014MNRAS.440..663Z, 2016ApJ...819..149A}). Very recently,  Ref.\cite{2017arXiv170505376R} claimed a $\ge4.5\sigma$ detection of a ring-like structure  at the outskirts of the clusters in the {\sl Fermi}-LAT stacking analysis of galaxy clusters.  Non-detection of gamma-rays from the central galaxy clusters thus poses  a challenge for the theoretical picture.

The Coma cluster of galaxies is the nearest massive clusters at a distance of $\sim 100 {\rm \ Mpc}$.  It shows evidence of efficient particle acceleration, as suggested by the presence of a giant radio halo and radio relics \cite{1993ApJ...406..399G,2003A&A...397...53T}.  The cluster lies near the north Galactic pole where the {Galactic} diffuse gamma-ray intensity is at a minimum. Together with the fact that there are broadband observations from radio to hard X-ray frequencies,  these make the Coma cluster a good candidate to search for CR-induced gamma-ray emission.  Ackermann et al. \cite{2016ApJ...819..149A} performed a {\em binned} likelihood analysis of the six years of {\it Fermi}-LAT Pass 8 data of the Coma cluster. They find two residual structures within the virial radius of the cluster, but the statistical significance of this emission is  below the threshold to claim detection of gamma-ray emission from the cluster. In this work, we perform an unbinned likelihood analysis of the nine years of {\it Fermi}-LAT Pass 8 data, with  careful modeling of the background sources. For the first time, we discover the gamma-ray emission above 200 MeV from the direction of the Coma cluster. We further find tentative evidence that the gamma-ray emission at the Coma center is spatially extended, which may be related to the bulk of the Coma  cluster.

\section {Data Selection And Analysis}
For this work, we use ~9 years (MET 239557417-522806178) of  public Pass 8 LAT data. We select FRONT+BACK converting photons corresponding to the SOURCE class with energies from 200 MeV to 300 GeV within a $12^{\circ}$ region of interest (ROI) centered at the Coma cluster center at $\alpha_{J2000}=194.95^{\circ},\delta_{J2000}=27.98^{\circ}$.
 We use recommended time selection of $\rm (DATA\_QUAL > 0) \&\& (LAT\_CONFIG == 1)$ and limit the data selection to zenith angles less than $90^{\circ}$, allowing us to effectively remove photons originating from the Earth limb.

\subsection{Background Model}
In our background model, we include all sources listed in the third {\it Fermi} catalog of point-like and extend sources (3FGL) \cite{2015ApJS..218...23A}, along with 4 non-3FGL point-like sources reported in Ref. \cite{2016ApJ...819..149A} within the region of ROI enlarged by $5^\circ$. We include also the standard diffuse emission background, i.e. the foreground for Galactic diffuse emission and the background for spatially isotropic diffuse emission \cite{2016ApJS..223...26A}, recommended for performing data analysis of Pass 8 LAT data. One shortcoming of using the 3FGL catalog (based on 4 yr of LAT observations) to search within a data set covering 9 yr is that spectral parameters may have substantially changed. To account for this variability, we leave the normalizations and spectral index free for all sources that are inside ROI and allow the normalization of the templates used to model the Galactic foreground and isotropic diffuse emission to vary freely. In addition, for the 3FGL sources outside ROI, we freeze all their parameters to the catalog values.
\begin{figure}[h]
\centering
\includegraphics[scale=0.4]{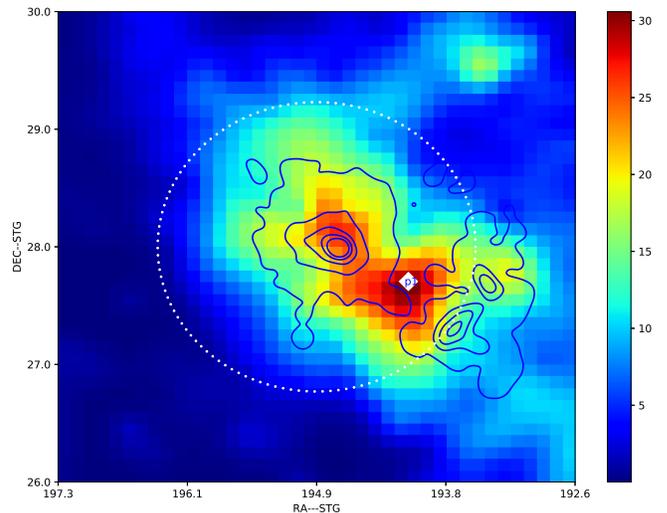}
\caption {Gaussian kernel ($\sigma = 0.1^{\circ}$) smoothed TS map of the Coma cluster region output from {\it gttsmap} in the energy band 0.2 - 300 GeV. The map has a dimension of $4^{\circ} \times 4^{\circ}$ and a resolution of  0.1$^{\circ}$ per pixel. The white dashed circle is the region subtended by the virial radius, $\theta_{200} = 1.23^{\circ}$. The diamond p1 represents the position of TS value peak. The contours correspond to measurements of the Coma cluster using the Westerbork Synthesis Telescope (WSRT) at a central frequency of 352 MHz (\cite{2011MNRAS.412....2B}). The WSRT observations are smoothed with gaussian kernel ($\sigma = 0.05^{\circ}$).  Contours start at 1 mJy beam$^{-1}$ and increase in steps of 6 mJy beam$^{-1}$.  }
\label{fig1}
\end{figure}
We make use of the Science Tools package (v10r0p5) with the  P8R2$\_$SOURCE$\_$V6 instrument response functions and carry out a standard unbinned likelihood analysis for the background. We use the {\it gttsmap} tool to search for any additional gamma-ray sources inside the Coma  cluster considering a $4^{\circ}\times 4^{\circ}$ region centered on the Coma center, see  \figurename~\ref{fig1}. For each pixel in the map the {\it gttsmap} tool evaluates the test statistic (TS), defined as ${\rm TS}=-2({\rm ln}L_0-{\rm ln}L)$, where $L_0$ is the maximum-likelihood value for null hypothesis and $L$ is the maximum likelihood with the additional source under consideration.
According to \figurename~\ref{fig1}, a residual structure appears to be roughly coincident with the radio halo region, but the position of peak TS values locates at the southwest side of the Coma cluster,  as shown by a diamond (namely p1) in \figurename~\ref{fig1}. The best position of p1 given by {\it gtfindsrc} tool is (194.148$^\circ$, 27.683$^{\circ}$) $\pm$ 0.093$^{\circ}$ (the uncertainty corresponds to $95\%$ confidence level).

After we submitted the first version of our paper {\footnote{We presented the discovery of the gamma-ray source in the first version of the submitted paper (arXiv:1709.08319v1), after which {\it Fermi}-LAT Collaboration distributed the preliminary {\it Fermi}-LAT list of sources (FL8Y). }, {\it Fermi}-LAT Collaboration distributed the preliminary {\it Fermi}-LAT list of sources (FL8Y) \footnote{https://fermi.gsfc.nasa.gov/ssc/data/access/lat/fl8y/} based on the first eight years Pass 8 data and the diffuse background of the 3FGL model. Our discovery of the excess at the position of p1 is confirmed by the FL8Y source list, with a name FL8Y J1256.6+2741.}  The {\it Fermi}-LAT Collaboration did not find any known source associated with FL8Y J1256.6+2741 using their automatic source association methods.
\begin{figure}[h]
\centering
\includegraphics[scale=0.4]{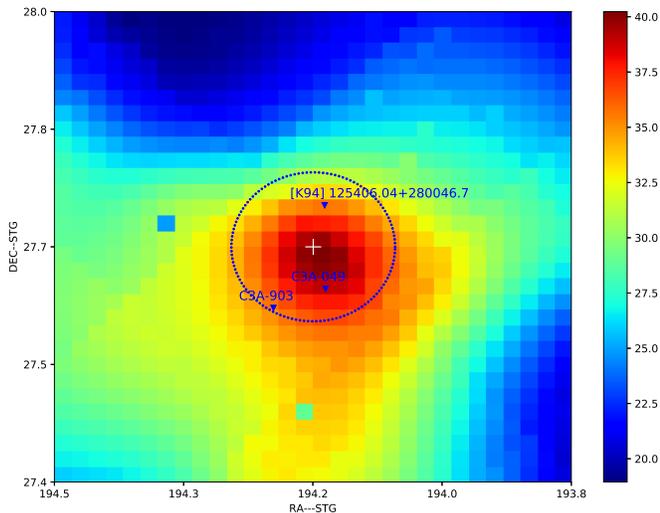}
\caption{$0.6^{\circ}\times 0.6^{\circ}$ TS map centered at the position of p1 in the energy band 0.2 - 300 GeV. The blue triangles represent the radio sources found in the catalogs of NED. The blue dashed circle is $95\%$ position uncertainty of p1 derived with the {\it gtfindsrc} tool. The white cross represents the position of p1. }
\label{figa7}
\end{figure}
 Since most of gamma-ray-emitting AGNs are sources of radio emission, we also search for possible radio sources associated with p1 in the radio source catalogs of NASA/IPAC EXTRAGALACTIC DATABASE (NED) {\footnote{https://ned.ipac.caltech.edu/forms/nearposn.html}}. Three faint radio sources are located within the $95\%$ containment radius given by the {\it gtfindsrc} tool, as shown in \figurename~\ref{figa7}. Their radio flux densities are $0.271\pm 0.096 \rm \ mJy$ for C3A-049, $0.290 \pm 0.096 \rm \ mJy$ for C3A-903 at 1.4 GHz  \cite{2009AJ....137.4436M}, and $11.2 \pm 1.4\rm \  mJy$ for [K94] 125406.04+280046.7 at 1.42 GHz \cite{1994A&AS..105..403K}.  We  can estimate the gamma-ray luminosity of these sources  if  the gamma-ray emission is mostly due to Compton scattering of the radio-producing electrons by the CMB, a reasonable expectation in light of the conclusion reached by Ref. \cite{2010Sci...328..725A} in their analysis of {\it Fermi}-LAT measurements of the radio galaxy Cen A. We find that the expected gamma-ray luminosity is so low that any physical association with the three radio sources is disfavored.

Since the number of sources inside ROI increases significantly  in the   FL8Y source list (compared to the 3FGL source list) , we use  the FL8Y source list as an updated background model in the subsequent analysis. Note that FL8Y J1256.6+2741 is treated as a source that we are interested.

\subsection{Cluster Spatial Modeling and Likelihood Analysis}
First, We assume that the gamma-ray excess within Coma cluster arises from a single component. We consider extended spatial templates similar to that in \cite{2016ApJ...819..149A}, which include 1) a cored profile template, and 2) a disk template. The predicted profile of the gamma-ray emission depends on the exact models and it could be shallower than the radio emission. The cored profile template in \cite{2016ApJ...819..149A} assumes that the profile of gamma-ray emission follows the relation $f_r (\theta)=(1+\frac{\theta^{2}}{r_c^{2}})^{-1/4}$, where $r_c=0.25^\circ$ is the core radius. The disk template assumes a uniform distribution of the gamma-ray emission up to the virial radius $R_{200}$ (here the subscript 200 refers to an enclosed density 200 times above the critical density of the Universe). Furthermore, assuming that the gamma-ray flux distribution traces the observed radio and X-ray flux distribution, we  consider two additional templates for the gamma ray emission profile, i.e., the radio emission template  and X-ray emission template. The radio emission template is based on the measurements of the Coma radio halo and relics using the Westerbork Synthesis Telescope (WSRT) at 352 MHz \cite{2011MNRAS.412....2B}. The X-ray emission template is based on  merged EPIC-pn image of the Coma cluster of galaxies with XMM-Newton in the 0.3-2.0 keV energy band \cite{2001A&A...365L..60B}. To investigate whether the gamma-ray emission within the Coma cluster could be attributed to a single point source, {we also consider the point source models A and B corresponding to point-like emission at the center of the Coma cluster and at the location of peak TS value (i.e., the position of p1),} respectively.

Since there are two residual structures, one being roughly coincident with the radio halo region and the other being coincident with the peak TS value, we also consider the two-component model (ext+p1), i.e., a spatially extended source plus a point-like source at the position of p1. To investigate whether  the additional source is extended, we also study the $\rm p_{center}+p1$ model, i.e., a point-like source at Coma center plus a point-like source at p1.
All these spatial models are shown in Table~\ref{tab1}.

To save the computation time, we perform an unbinned likelihood analysis fixing the parameters of all point-like sources to their maximum likelihood values obtained from the background-only model fit. The normalizations of the Galactic and isotropic diffuse components are still left free.

\begin{table*}
\centering
\begin{threeparttable}
\caption{ Unbinned likelihood analysis results for energy band $200 \rm\ MeV-300\rm\ GeV$}
\begin{tabular}{lcccr}
\hline
\hline
Spatial Model & Photon Flux &Energy Flux&  Power-law Index & TS \\
 &($\times 10^{-9}{\rm \ ph\ cm^{-2}\ s^{-1}}$)&($\times 10^{-12}{\rm \ erg\ cm^{-2}\ s^{-1}}$) & & \\
 \hline

Disk & 3.14 $\pm$ 0.54 & 2.52 $\pm$ 0.59 & 2.65 $\pm$ 0.25 & 38.9\\
Core & 3.08 $\pm$ 0.52 & 2.50 $\pm$ 0.59 & 2.64 $\pm$ 0.25 & 40.1\\
Radio & 2.74 $\pm$ 0.48 & 2.11 $\pm$ 0.43 & 2.70 $\pm$ 0.24 & 42.9\\
X-ray & 2.39 $\pm$ 0.44 & 1.70 $\pm$ 0.35 & 2.81 $\pm$ 0.28 & 37.2\\
Point Source A ($\rm p_{center}$) & 1.94 $\pm$ 0.42 & 1.12 $\pm$ 0.43 & 3.24 $\pm$ 0.94 & 23.4\\
Point Source B (p1)& 1.92 $\pm$ 0.43 & 1.45 $\pm$ 0.26 & 2.73 $\pm$ 0.19 & 41.3\\
\hline
Disk+p1 & 2.45 $\pm$ 0.65 & 1.78 $\pm$ 0.81 & 2.78 $\pm$ 0.53 & 53.4\\
   & 0.67 $\pm$ 0.35 & 0.82 $\pm$ 0.31 & 2.30 $\pm$ 0.26 &  \\
\hline
Core+p1 & 2.43 $\pm$ 0.63 & 1.82 $\pm$ 0.76 & 2.73 $\pm$ 0.46 & 54.3\\
 & 0.65 $\pm$ 0.35 & 0.80 $\pm$ 0.30 & 2.30 $\pm$ 0.27 &  \\
\hline
Radio+p1 & 2.25 $\pm$ 0.55 & 1.66 $\pm$ 0.49 & 2.76 $\pm$ 0.36 & 56.5\\
 & 0.53 $\pm$ 0.30 & 0.75 $\pm$ 0.30 & 2.22 $\pm$ 0.27 &  \\
\hline
X-ray+p1 & 1.79 $\pm$ 0.53 & 1.23 $\pm$ 0.44 & 2.81 $\pm$ 0.28 & 52.9\\
 & 0.72 $\pm$ 0.37 & 0.85 $\pm$ 0.29 & 2.33 $\pm$ 0.27 &  \\
\hline
p$\rm _{center}$+p1 & 1.13 $\pm$ 0.51 & 0.65$\pm$ 1.08 & 3.23 $\pm$ 4.00 & 45.7\\
 & 1.14$\pm$ 0.42 & 1.08 $\pm$ 0.38& 2.49$\pm$ 0.27 &  \\
\hline
\hline
\end{tabular}
\begin{tablenotes}
\item\textbf{Notes.} The Disk model is a uniform disk with a radius corresponding to the virial radius  $\theta_{200}$. The Core model is a predicted gamma-ray flux profile (see the text for details). The point source model A and B correspond to point sources at the center of the Coma cluster (p$\rm_{center}$) and at the position of p1 in \figurename~\ref{fig1}, respectively. In each two-component model,  the flux and spectral index of each single component are listed in the top/bottom line, which the bottom line corresponds to point source at p1. The associated uncertainty refers to the 68\% error reported by HESSE algorithm embedded in the {\it gtlike} tool.
\end{tablenotes}
\label{tab1}
\end{threeparttable}
\end{table*}

\section{Results}
The results for the unbinned likelihood analysis are summarized in Table~\ref{tab1}. Comparing the TS values of various single-component models, we reject the hypothesis that all the gamma-ray emission originates from the centre point-like source because of the significantly lower TS value for the point source model A \cite{2012ApJ...756....5L}.  However, we can not distinguish between the point source model B and the single-component extended emission models, since their TS values are close.

Further, by comparing the two-component models with the point source model B, we investigate weather the  gamma-ray emission within Coma cluster entirely arises from the possible point-like source at p1. We find that the TS value is increased by 15.2 for the radio + p1 model compared
to the point source model B, supporting the presence of an additional gamma-ray source. To evaluate the false detection probability for an additional source, we performed 800 simulations, assuming the point source model B is the true source model of emission within the Coma cluster
and fitting the simulated data with the radio + p1 model.  It is found that only 1 out 800 simulations result in $\rm TS_{radio+p1}-TS_{p1}>15.2$, corresponding to a chance occurrence of $< 1.25\times 10^{-3}$ ($> 3.0 \sigma$) (see  Appendix A for more details).

\begin{figure}
\centering
\includegraphics[scale=0.4]{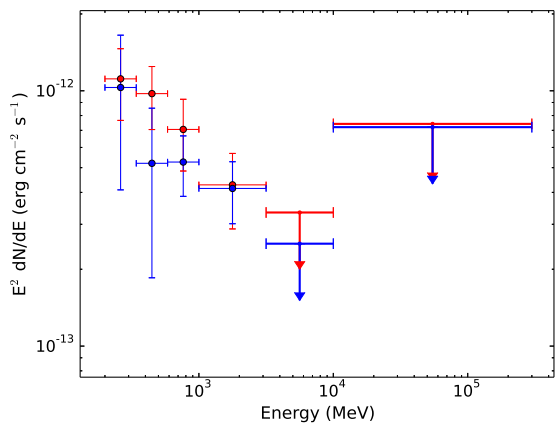}
\caption{{ Spectral energy distributions (SEDs) of the extended emission component of the Coma cluster for the single radio model (red) and $\rm radio+p1$ model (blue).} The upper limits at 95\% confidence level are derived when the TS value for the data points are lower than 4.}
\label{fig2}
\end{figure}

In addition, we investigate the spatial extension of the additional source at the Coma center by comparing the $\rm radio+p1$ model with $\rm p_{center}+p1$ model. We find the TS value is increased by 10.8 (i.e., $\rm TS_{radio+p1}-TS_{p_{center}+p1}=10.8$ ), which favors that the source at the Coma center is  extended. For the two-component models, we find that the additional extended component always contributes a dominant part to the whole gamma-ray flux. We  find that the  the gamma-ray spectrum of the extended component is soft, with a photon spectral index of $\Gamma=2.6-2.8$ ($dN_\gamma/d\epsilon_\gamma\propto\epsilon_\gamma^{-\Gamma}$). Considering the extended radio templates, we provide the spectral energy distribution (SED) for the extended component,  which is showed in \figurename~\ref{fig2}. The energy flux in $0.2-300\rm\ GeV$ is about $2\times 10^{-12} {\rm\ erg\ cm^{-2}\ s^{-1}}$, leading to a total gamma-ray luminosity of $L_\gamma({\rm 0.2-300\ GeV})\simeq 2\times10^{42}{\rm\ erg\ s^{-1}}$.

\section{Systematic Uncertainties}
To assess the robustness of our results we perform a number of systematic checks. {Concerning the favored $\rm radio+p1$ model, we quantify the variation of the photon flux and spectral index for the extended component. In particular, we investigate how  the detection significance of the gamma-ray emission ($\rm TS_{radio+p1}$), the significance  for the presence of an additional source at the Coma center (indicated by $\rm TS_{radio+p1}-TS_{p1}$), and the significance for an extended spatial structure (indicated by  $\rm TS_{radio+p1}-TS_{p_{center}+p1}$) change in the systematic checks.
\subsection{ROI Size and Free Sources}
 We first vary the ROI size from $8^\circ$ to $15^\circ$ radius to repeat our analysis and find that the photon flux and spectral index change by at most  $\sim 5\%$. The variations are at most 4 for the $\rm TS_{radio+p1}-TS_{p1}$ and less than 2  for $\rm TS_{radio+p1}-TS_{p_{center}+p1}$, indicating a small impact.
Similarly, we vary the radius within which the spectrum parameter of point-like sources are free. The changes of the resulting photon flux and spectral index are at most $3\%$.  {The impacts on the detection significance and the significance for an extended spatial structure  are both small.}
\subsection{Event Classes}
At a high galactic latitude where Coma is located, the residual cosmic rays are non-negligible  contamination, especially for analyzing a source with relatively large extension. We therefore repeat our analysis using the CLEAN and ULTRACLEANVETO event class data\footnote{https://fermi.gsfc.nasa.gov/ssc/data/analysis/documentation/Cicerone/Cicerone\_Data/LAT\_DP.html}, which have lower background rates but smaller effective areas. To check the impact due to the angular resolution of the data set, we also do an analysis for the front conversion events of SOURCE event class, which have intrinsically better angular resolution (also smaller effective areas).  We  find  that using the data sets with different event class or using only the front conversion data  change the photon flux and the spectral index by at most $18\%$. Due to decreasing event counts by $\sim 40\%$ for ULTRACLEANVETO class data and FRONT event type data, the decrease of the $\rm TS_{radio+p1}$ value of $\sim 40\%$ is reasonable, as shown in \figurename~\ref{figa3}. We also find that the variations of the $\rm TS_{radio+p1}-TS_{p1}$ and $\rm TS_{radio+p1}-TS_{p_{center}+p1}$ values, compared to the standard result listed in Table I, are less than 4 in all the tests, implying a small impact due to selection of event class.
\begin{figure}[h]
\centering
\includegraphics[scale=0.4]{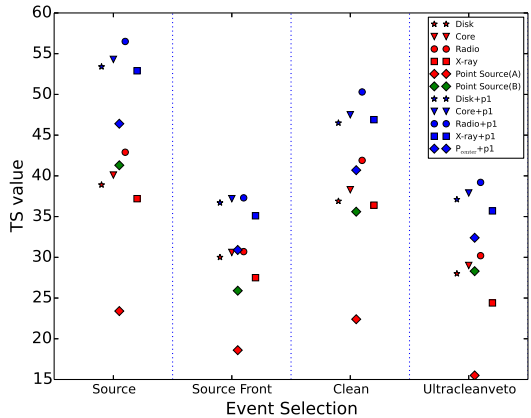}
\caption{TS value of the Coma emission for various spatial models using different event class data. }
\label{figa3}
\end{figure}
\subsection{Low Energy Thresholds }
We study the impact caused by different low energy thresholds of  100 MeV, 300 MeV, 500 MeV and 800 MeV. The spectral index changes by less than 8\%. We find that the detection significance of the source decreases as the low energy threshold increases, which is simply due to decreasing number of event counts. However, the detection significance for an additional source and the  significance for the extended emission, as indicated by $\rm TS_{ext+p1} - TS_{p1}$ and  $\rm TS_{radio+p1}-TS_{p_{center}+p1}$ respectively, roughly keeps the same for different low energy thresholds (see \figurename~\ref{figa4}). It is noteworthy that the bright source 3FGL J1224.9+2122 (namely FL8Y J1224.9+2122 in FL8Y source list) can affect the TS value  for 100 MeV - 300 GeV energy band analysis, although it is located $\sim 10.5^\circ$ away from Coma center.
 If we fix the spectral parameters of this source to those in 3FGL, the obtained results of the extended Coma emission would differ from the case when its spectral parameters are left free.
 We find the impact from the bright source can be neglected when we increase the lower energy threshold to $\ge 200$ MeV or fix its parameters to those reported in the FL8Y source list.

\begin{figure}[h]
\centering
\includegraphics[scale=0.4]{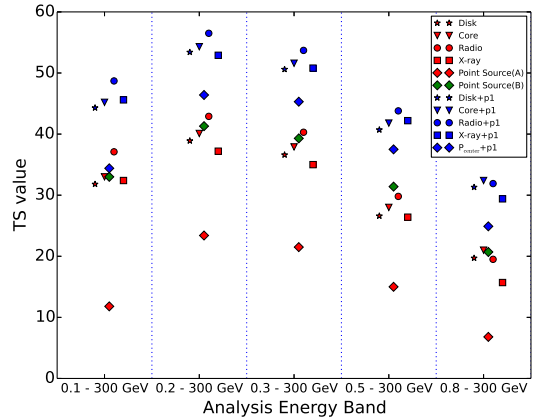}
\caption{The TS values of the Coma emission for various spatial models using different low-energy thresholds.}
\label{figa4}
\end{figure}

\subsection{Diffuse background models}
Although Coma is located at a high galactic latitude, the uncertainty from Galactic diffuse foreground modeling may still be significant. We thus compare results obtained by using the standard diffuse emission model with those obtained by  using alternative diffuse emission models.
We use the maps of the predicted Galactic diffuse gamma-ray emission derived in Ref. \cite{2012ApJ...750....3A} as the start point of our template creation \footnote{https://galprop.stanford.edu/PaperIISuppMaterial/} . Ref. \cite{2012ApJ...750....3A} provides 128 sets of maps corresponding to different model parameters.  We adopt 16 sets among them, which varies in the most important parameters involved in the template creation, including CR source distribution (Lorimer, SNR), halo size (4 kpc, 10 kpc), spin temperature (150 K, $10^5$ K), and $E(B - V)$ magnitude cut (2 mag, 5 mag) \citep{2012ApJ...750....3A,2013arXiv1304.1395D}. We use the spatial templates for $\pi^0$ decay, bremsstrahlung radiation and IC gamma-rays generated by GALPROP \cite{2011CoPhC.182.1156V} to replace the standard Galactic foreground model. We also use a different set of isotropic diffuse templates that are created to accommodate the alternative Galactic diffuse models. For simplicity, we do not fit the different components along the line of sight separately for these models, but only adopted a free overall normalization for each of the $\pi^0$, bremsstrahlung and IC templates, since at high Galactic latitudes the vast majority of gas resides in the neighborhood of the solar system.  We emphasize that these 16 models may not cover the complete uncertainty of the systematics involved in interstellar emission modeling. The resulting uncertainty should therefore only be considered as an indicator of the systematic uncertainty due to the mis-modeling of the Galactic diffuse foreground emission.  We find that varying the diffuse emission models changes  the photon flux and spectral index by at most $\sim 30\%$. As shown in \figurename~\ref{figa5}, the value of $\rm TS_{radio+p1}-TS_{p1}$ ranges from 15 to 21, indicating a small impact for the detection significance of an additional source. In addition, the value of $\rm TS_{radio+p1}-TS_{p_{center}+p1}$ is in the range of 10-17 , which implies that different diffuse models do not  change the nature of the extendedness of the additional gamma-ray emission.
\begin{figure}[h]
\centering
\includegraphics[scale=0.4]{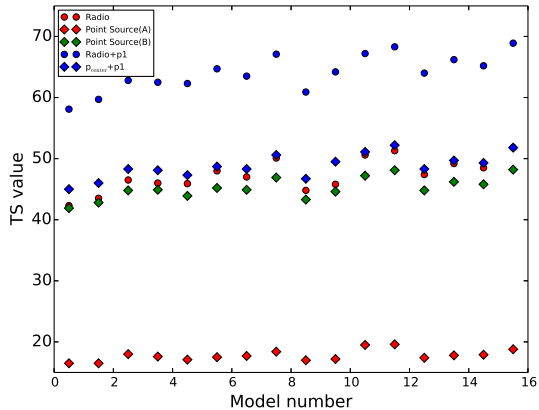}
\caption{The TS values of Coma emission derived by an unbinned analysis for 16 alternative Galactic diffuse gamma-ray templates.}
\label{figa5}
\end{figure}

We summarize the results on systematic uncertainties  in Table~\ref{tab2}. The largest uncertainty in the photon flux and spectral index arises from the uncertainty in the Galactic diffuse foreground, while  other discussed systematic effects have only minor influence.  }

\begin{table*}
\centering
\begin{threeparttable}
\caption{Systematic Uncertainties}
\begin{tabular}{lcccccc}
\hline
\hline
Type & Variation of & Photon Flux & Spectral Index & Range of & Range of & Range of $\rm TS_{radio+p1}-$ \\
& Input Parameters& Impact & Impact &  $\rm TS_{radio+p1}$ & $\rm TS_{radio+p1}-\rm TS_{p1}$ & $ \rm TS_{p_{center}+p1}$ \\
\hline

ROI size \tablenotemark{} & $ 8^\circ -15^\circ$ & $ <5\%$ & $ <3\%$ & (56 , 60) &(12 , 17)&(9 , 12)\\
Free radius\tablenotemark{} & $4^\circ -12^\circ$ & $ <3\%$ & $ <3\%$& (55 , 58)& (14 , 17)&(10 , 12)\\
Event class & alt. event class \tablenotemark{} & $ <16\%$ & $ <7\%$& (37 , 57) & (11 , 16)&(6 , 11)\\
Low energy threshold \tablenotemark{}& 100 - 800 MeV & $ -- $& $ <8\%$ & (33 , 57)& (11 , 16)&(6 , 14) \\
Diffuse modeling& alt. diffuse models \tablenotemark{} & $ <30\%$ & $ <18\%$& (56 , 69)& (15 , 21)&(10 , 17)\\

\hline
\hline
\end{tabular}
\begin{tablenotes}
\item\textbf{Notes.} Overview of systematic uncertainties.
\item[a]  We select the ROI size of $8^\circ$ , $10^\circ$, and $15^\circ$ radius to repeat our analysis.
\item[b]  We fix the normalizations and spectral indexes to the background-only fitting values for sources outside the region of $4^\circ$, $6^\circ$, $8^\circ$ and $12^\circ$.
\item[c] The CLEAN, UNTRACLEANVETO, and FRONT conversion SOURCE events are used to perform the analysis.
\item[d] The low energy thresholds of 100, 300, 500 and 800 MeV are used.
\item[e] See the Part D of Section 4 for details.
\end{tablenotes}
\label{tab2}
\end{threeparttable}
\end{table*}

\section{Discussions}
Compared with Ref.\cite{2016ApJ...819..149A}, our analysis is different in the following aspects:
1) We use an unbinned likelihood analysis, which is useful for faint sources although it costs more computing time;
2) We use an updated point-like source list (i.e., FL8Y) and free more background source model parameters;
3) We use a larger data set of 9 years of {\it Fermi}-LAT observations;
4) We consider more spatial models, especially the two-component models, which are not discussed in earlier works.
 Our result constitutes the first detection of GeV emission from the direction of the Coma cluster. We also find tentative evidence that the gamma-ray emission at the Coma center is spatially extended.

To compare our results with that in Ref.\cite{2016ApJ...819..149A}, we also perform a binned likelihood analysis with events selection similar to the unbinned likelihood analysis. We  find that the improvement of TS value for an additional source, as indicated by $\rm TS_{radio+p1}-TS_{p1}$,  ranges from 6 to 12,  which is slightly lower than that derived by the unbinned analysis. The  significance for the extended gamma-ray emission, as indicated by ($\rm TS_{raido+p1}-TS_{p_{center}+p1}$),  is lower than that given by the unbinned analysis (see the Appendix B for more details).

It has been suggested that relativistic electrons can produce gamma-rays through inverse-Compton (IC) scattering of CMB photons \cite{2000Natur.405..156L,2001ApJ...562..233M,2007IJMPA..22..681B,2009JCAP...08..002K,2010MNRAS.409..449P}, while protons produce gamma-rays through $pp$ collisions with the ICM.  Theoretically, it is expected that gamma-ray emission above  200 MeV is dominated by the hadronic process in the cluster core region, whereas in the outskirts region the IC emission is the dominant component \cite{2001ApJ...562..233M,2010MNRAS.409..449P}.  Assuming a single power-law spectral extrapolation,  a bright hard X-ray emission would be expected due to the soft spectrum with an photon index of $\Gamma\simeq2.7$. However, a recent joint analysis of {\em Swift} Burst Alert Telescope and the XMM-Newton observations places a conservative limit on the non-thermal, hard X-ray emission of $\le 4.2\times10^{-12}{\rm\ erg\ cm^{-2}\ s^{-1}}$ in 20 - 80 keV \cite{2011ApJ...727..119W}, which disfavors the IC process as the dominant mechanism to produce the gamma-ray emission. Of course, the assumption of a single power-law electron spectrum  may not be true as the acceleration mechanism is not  well-understood for cluster shocks, so we can not  rule out the leptonic scenario.  On the other hand, the hadronic scenario, invoking the cosmic-ray proton interaction, does not suffer from this spectrum constraint due to the characteristic pion-decay gamma-ray spectrum.  The flux of gamma-rays produced by cosmic-ray protons can be used to infer the cosmic ray content in ICM.  This is reflected by the volume-averaged value of the CR to thermal energy ratio, defined as $f_{\rm cr}=\frac{U_{\rm cr}}{U_{\rm th}}$, where $U_{\rm cr}$ is the  CR energy density and $U_{\rm th}=\frac{3}{2} n_g k T$ is the thermal energy density. By comparing the observed gamma-ray flux with the prediction, we find  $f_{\rm cr}\sim2\%$ for the Coma cluster (see the Appendix C for more details).

Beside the structure that are roughly coincident with the radio halo, another residual structure is evident on the southwest side of the radio halo. This structure is spatially coincident with a sharp, low surface-brightness front of synchrotron radio emission (see {Figure 3} of Ref.\cite{2011MNRAS.412....2B}). There is also a corresponding edge in the X-ray surface-brightness and jump in temperature (see Fig.2 of Ref.\cite{2003A&A...400..811N}). This structure may reflect a shock induced by the infall of a substructure onto the Coma cluster \cite{2003A&A...400..811N,2011MNRAS.412....2B}. The gamma-ray emission around this structure may not be produced dominantly by the hadronic process, since the density of the ICM, as the target for $pp$ collisions, is low at this radius. Then the gamma-ray emission should arise from the IC process, which indicates that the shock must be able to accelerate electrons to energies of at least $\gamma_e\sim10^6$.

\acknowledgments
We thank L. Rudnick for providing us the radio data of the Coma cluster that we used for  the radio emission template. We also thank G. Brunetti,  Z.-X. Wang and T. Tam for useful discussions. This work has made use of data and software provided by the
Fermi Science Support Center. This work is supported by the 973
program under grant 2014CB845800, the NSFC under grant
11625312 and 11851304.

\appendix

\section {Simulation}
\begin{figure}[h]
\centering
\includegraphics[scale=0.4]{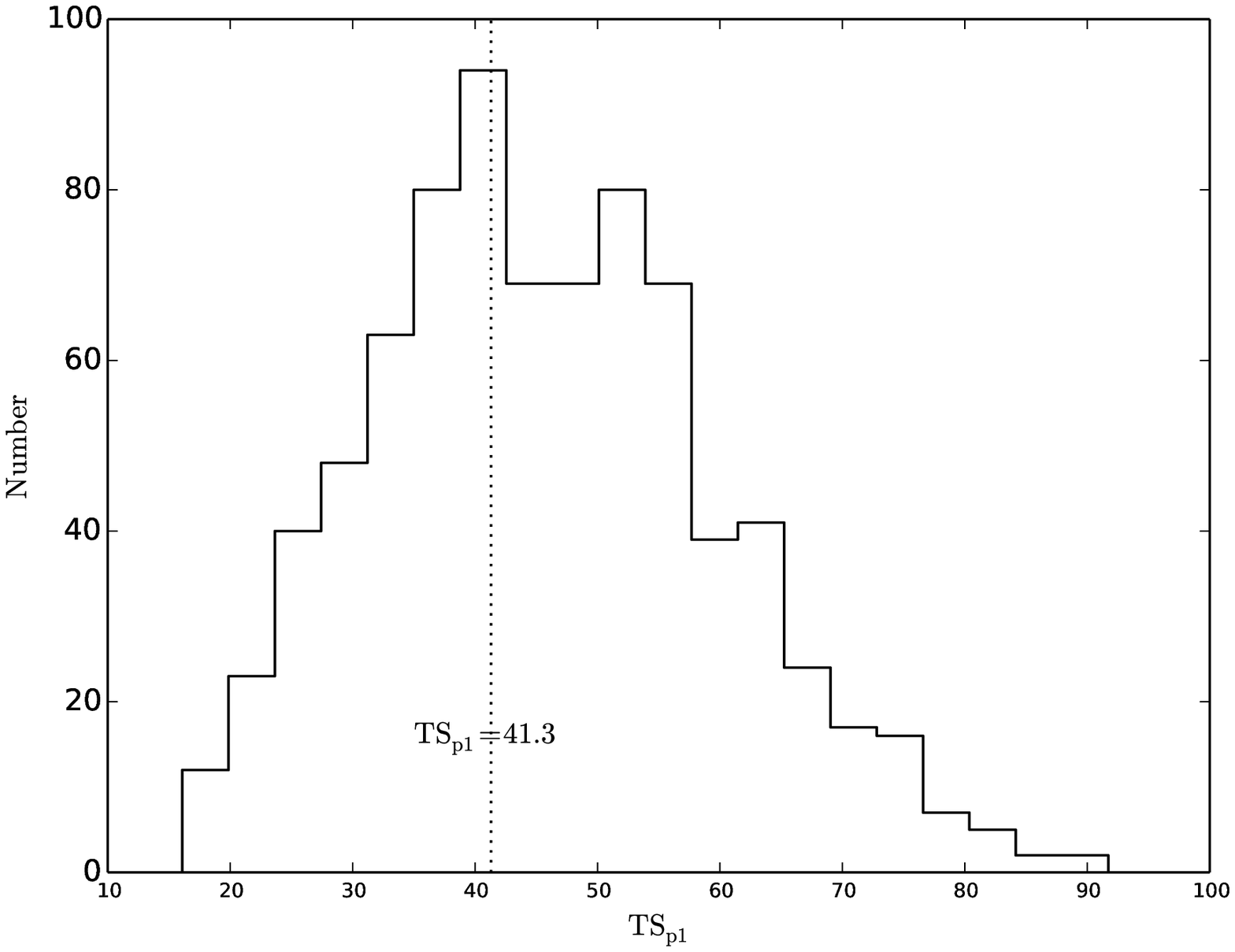}\\
\includegraphics[scale=0.4]{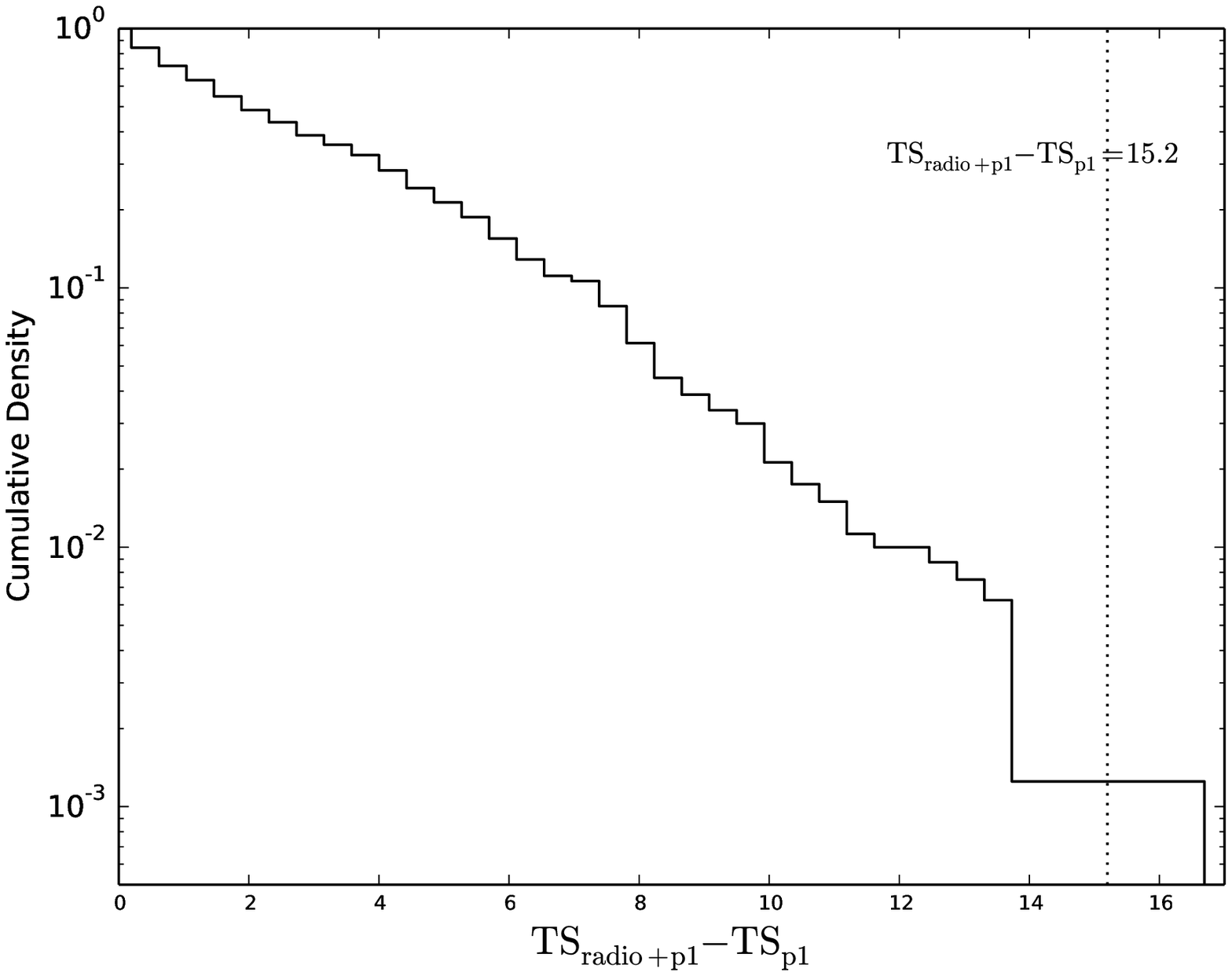}
\caption{Top: Distribution of TS values for a point-like source at p1 derived from fitting 800 Monte Carlo data sets with the point source model B. Bottom: Cumulative distribution of the ${\rm TS_{radio+p1}-TS_{p1}}$ for estimating the detection significance of the additional emission component.}
\label{figa1}
\end{figure}

We first simulate the background, including the sources listed in FL8Y source list, Galactic diffuse emission and isotropic diffuse emission, using the LAT simulation tool {\it gtobssim}. The background sources are simulated with the same spectral and spatial model parameters as our results of the background-only fit. To evaluate the detection significance of an additional source assuming a point-like source at the position of p1, we further simulate the point-like source of p1 on top of the simulated background. The point-like source is simulated with a power-law
spectral model with integrated flux of $1.92 \times 10^{-9} \rm \ ph\ cm^{-2}\ s^{-1}$ in the energy range from 200 MeV to 300 GeV and a photon spectral index of 2.74, which are the best-fit parameters for the point source model B. In total, 800 Monte Carlo data sets are generated. We then do an unbinned likelihood analyses for these data sets using the point source model B and the $\rm radio+p1$ model respectively. Considering the large amount of computational time for unbinned analyses,  we fix all the background parameters in the fit to the simulated data, but free the normalizations of the Galactic and isotropic diffuse background. The distributions of $\rm TS_{p1}$ and $\rm TS_{radio+p1}-TS_{p1}$  are plotted in  \figurename~\ref{figa1}.  Only 1 out 800 simulations result in $\rm TS_{radio+p1}-TS_{p1}>15.2$, corresponding to a chance occurrence of $< 1.25\times 10^{-3}$, i.e., the detection significance of $> 3.0 \sigma$ confidence level.

\begin{figure}[h]
\centering
\includegraphics[scale=0.4]{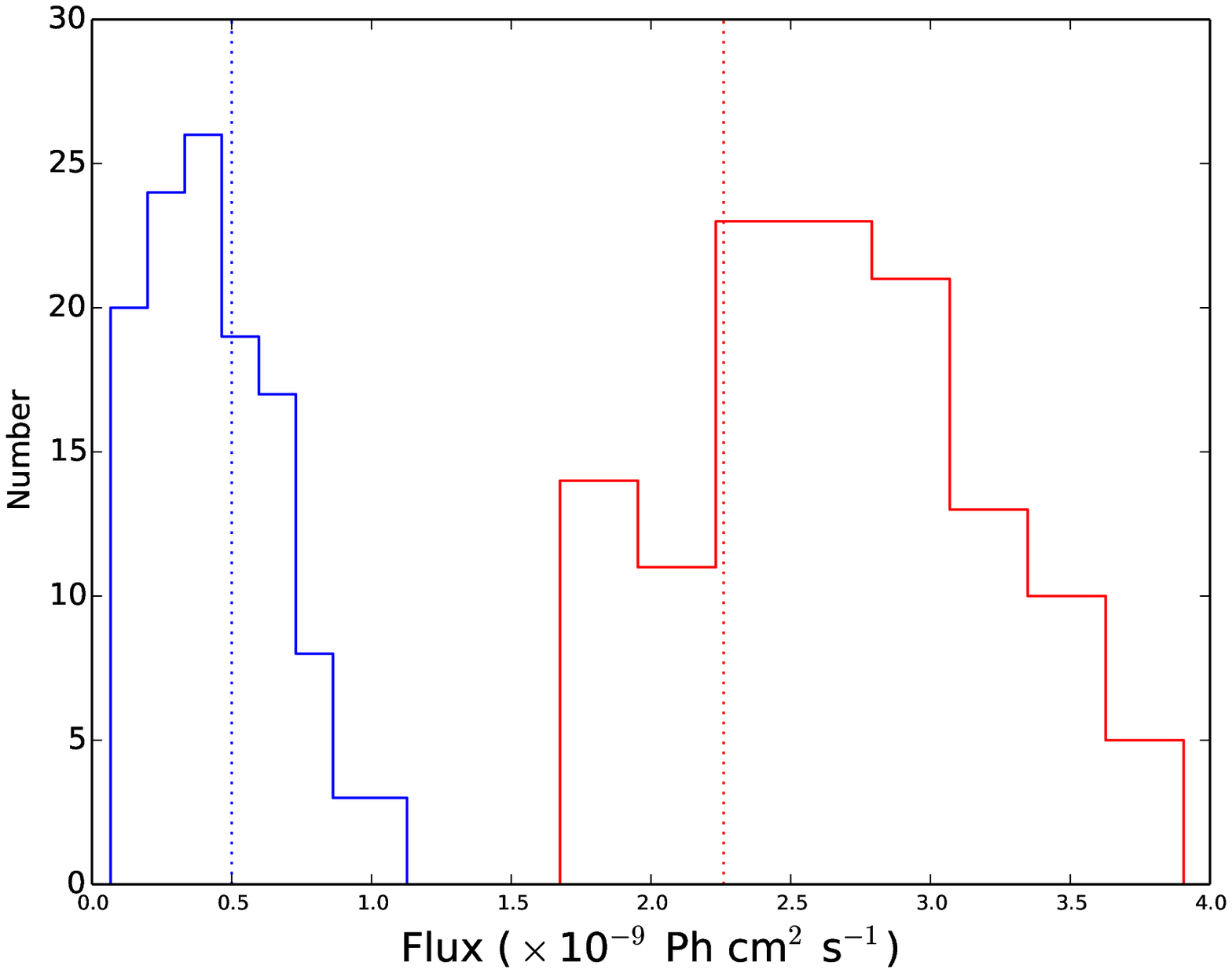}\\
\includegraphics[scale=0.4]{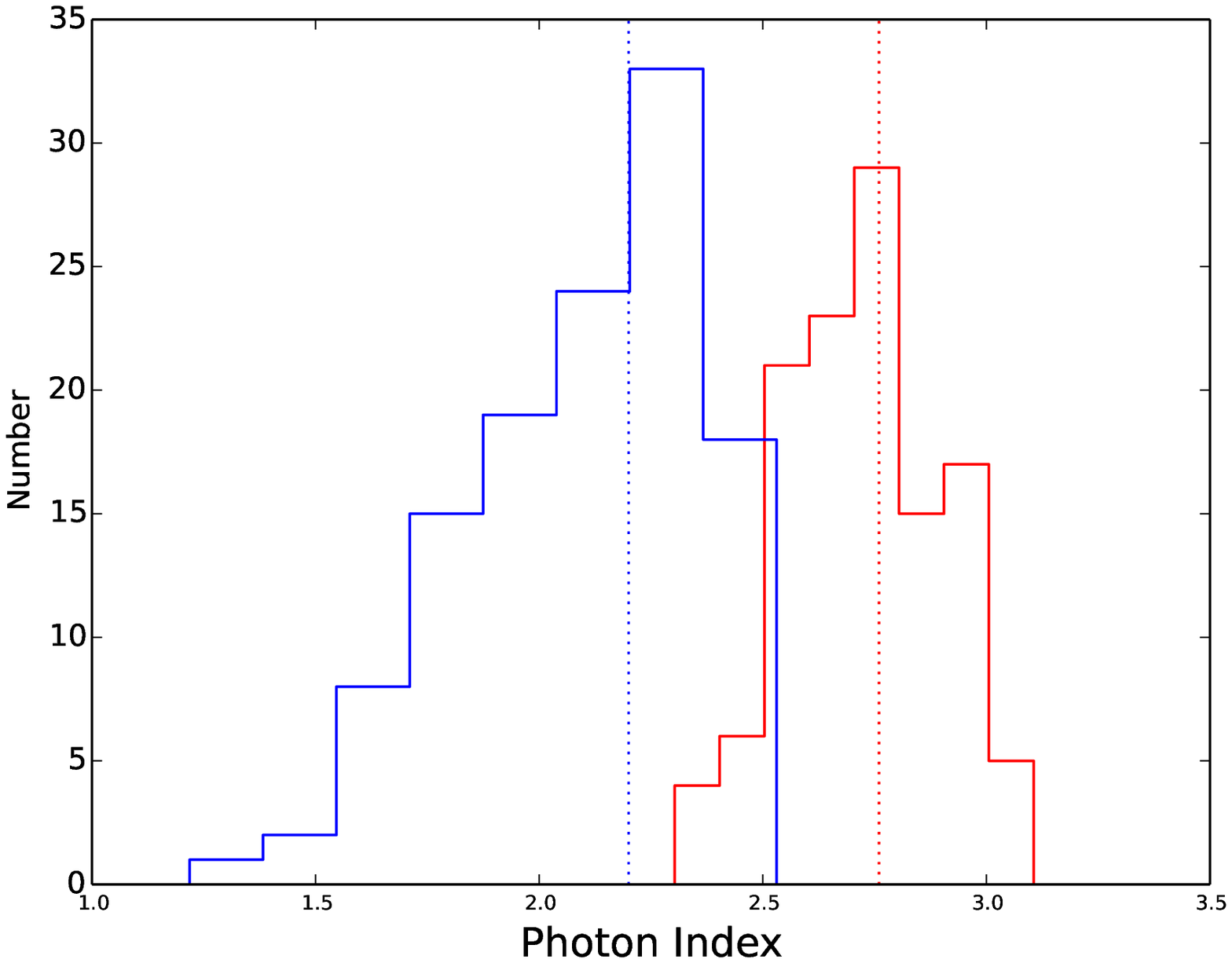}
\caption{Distribution of the photon flux (top) and photon index (bottom) of the respective radio and p1 components derived from fitting 120 simulated data sets with the $\rm radio+p1$ model. The red and blue are for the radio emission component and the point-like emission component, respectively. The dashed lines represent the input  parameters of the simulations.}
\label{figa2}
\end{figure}

We perform a second set of 120 simulations to survey the impact of parameter degeneracy between the extended emission and the point source emission. We simulate a source with the  same spatial distribution and spectral parameters (Table I)  as that obtained in the $\rm radio+p1$ model fit on top of the background, and then do an unbinned likelihood analysis using the $\rm radio+p1$ model. As shown in  \figurename~\ref{figa2}, we  find a small bias of $\sim 0.4 \times 10^{-9}{\rm \ ph\ cm^{-2}\ s^{-1}}$ in flux for the radio component compared with our input flux ($\sim 2.2 \times 10^{-9}{\rm \ ph\ cm^{-2}\ s^{-1}}$), indicating only a small degeneracy between the extended emission and the point source emission.

\section {A binned likelihood analysis }
\begin{figure}[h]
\centering
\includegraphics[scale=0.4]{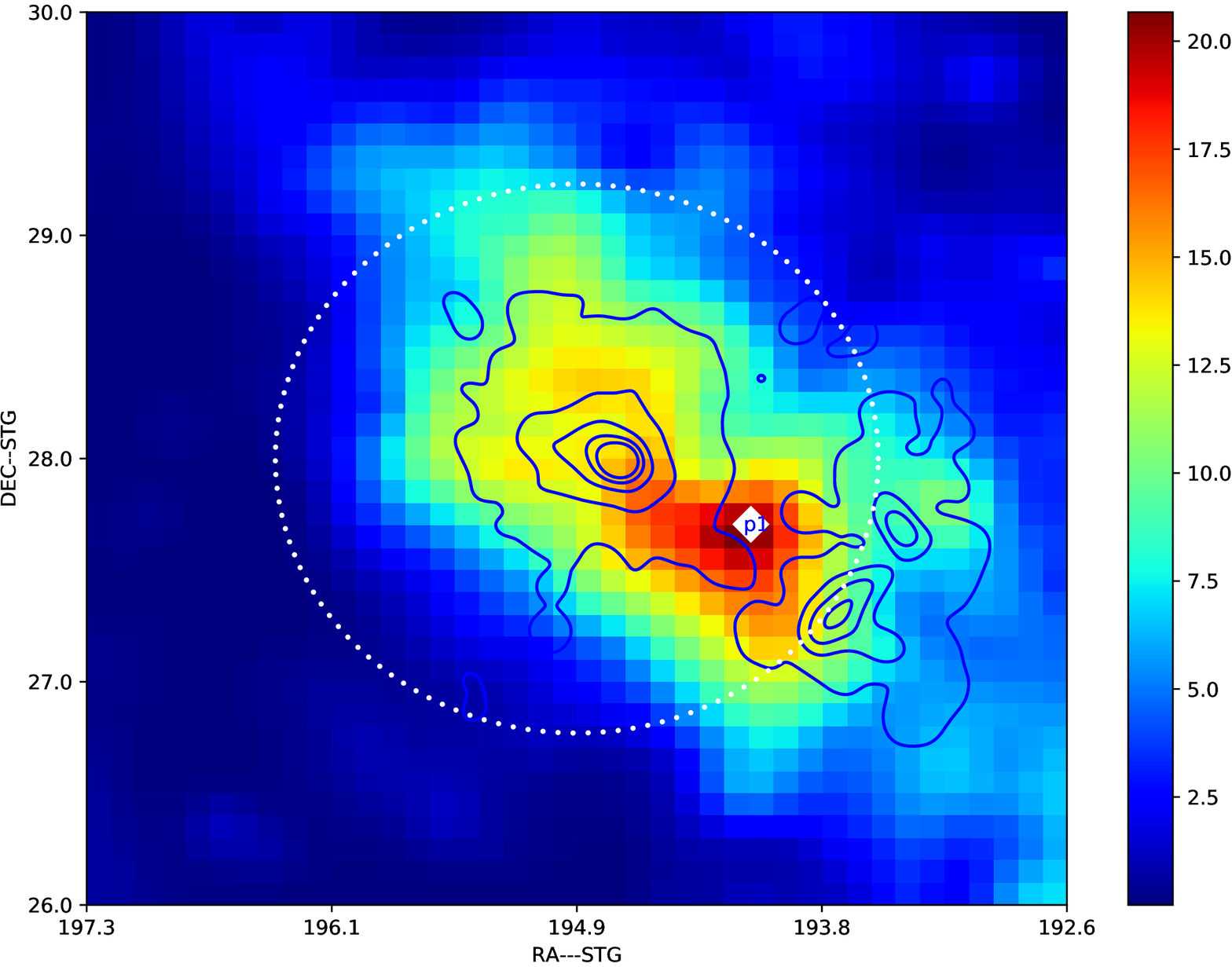}
\caption{Gaussian kernel ($\sigma = 0.1^{\circ}$) smoothed TS map for the binned analysis in the energy band 0.2 - 300 GeV.  The denotations of the white dashed circle, diamond and blue counter are the same as Fig. 1.}
\label{figa6}
\end{figure}

To compare our results with that in Ref.\cite{2016ApJ...819..149A}, we also perform a binned likelihood analysis with event selection similar to the unbinned likelihood analysis.  We select photons with energies from 200 MeV to 300 GeV within a square of $15^{\circ} \times 15^{\circ}$  centered at the Coma cluster center. Similar to the unbinned analysis, we use the same data cut, including zenith angles less than 90$^{\circ}$ and time intervals of $\rm (DATA\_QUAL > 0) \&\& (LAT\_CONFIG == 1)$. We divide our data into 24 logarithmically spaced bins in energy and use a spatial binning of 0.1$^{\circ}$ per pixel. We perform a background-only fit and adopt the same free parameter strategy as that of the unbinned analysis, as described in Part A of Section 2. The {\it gttsmap} tool produces a TS map with a dimension of $4^{\circ} \times 4^{\circ}$,  shown in \figurename~\ref{figa6}, in which  similar residual structures are found.

\begin{figure}[h]
\centering
\includegraphics[scale=0.4]{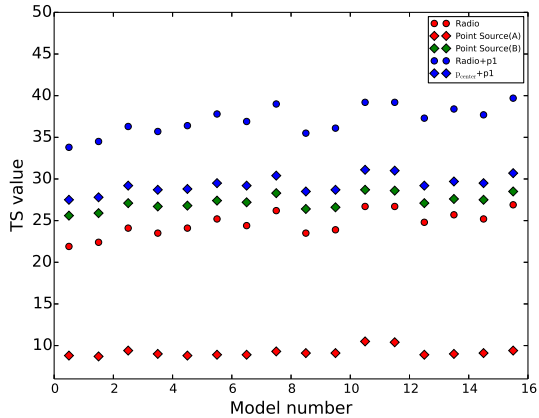}
\caption{The TS values of Coma emission derived by a binned method for 16 alternative Galactic diffuse gamma-ray templates.}
\label{figa8}
\end{figure}

Considering the spatial templates listed in Table I, we perform a binned likelihood analysis for the gamma-ray emission within the Coma cluster. Due to that the largest system uncertainty originate from the diffuses emission models, we also perform a test using 16 alternative diffuse emission models.  The results are shown in Table~\ref{taba1} and \figurename~\ref{figa8}. We note that the significance of the detection decreases obviously, with $14 < {\rm TS} < 32$ for different spatial models. We  find that the improvement of TS value for an additional source, as indicated by $\rm TS_{radio+p1}-TS_{p1}$,  ranges from 6 to 12,  which is slightly lower than that derived by the unbinned analysis. The  significance for the extended gamma-ray emission, as indicated by ($\rm TS_{raido+p1}-TS_{p_{center}+p1}$),  is also lower than that given by the unbinned analysis.

{We check the TS values for the binned analysis using  a smaller spatial bin (i.e., a  binning of 0.025 $^{\circ}$ per pixel). We find that the significance has only a very small increase (the TS value increases by only  2 ).}

\begin{table*}
\centering
\begin{threeparttable}
\caption{ Binned likelihood analysis results for energy band $200 \rm\ MeV-300\rm\ GeV$}
\begin{tabular}{lcccr}
\hline
\hline
Spatial Model & Photon Flux &Energy Flux&  Power-law Index & TS \\
 &($\times 10^{-9}{\rm \ ph\ cm^{-2}\ s^{-1}}$)&($\times 10^{-12}{\rm \ erg\ cm^{-2}\ s^{-1}}$) & & \\
 \hline

Disk & 2.15 $\pm$ 0.57 & 1.47 $\pm$ 0.65 & 2.88 $\pm$ 0.71 & 15.4\\
Core & 2.13 $\pm$ 0.56 & 1.53 $\pm$ 0.59 & 2.80 $\pm$ 0.54 & 16.4\\
Radio & 2.01 $\pm$ 0.50 & 1.55 $\pm$ 0.41 & 2.70 $\pm$ 0.33 & 21.5\\
X-ray & 1.79 $\pm$ 0.47 & 1.27 $\pm$ 0.34 & 2.81 $\pm$ 0.36 & 19.4\\
Point Source A & 1.65 $\pm$ 0.50 & 0.91 $\pm$ 0.57 & 3.39 $\pm$ 2.31 & 14.3\\
Point Source B & 1.49 $\pm$ 0.44 & 1.18 $\pm$ 0.26 & 2.67 $\pm$ 0.25 & 25.3\\
\hline
Disk+p1 & 1.15 $\pm$ 0.68 & 0.48 $\pm$ 0.41 & 5.14 $\pm$ 4.12 & 28.3\\
 & 0.93 $\pm$ 0.37 & 0.97 $\pm$ 0.26 & 2.41 $\pm$ 0.22 &  \\
\hline
Core+p1 & 1.43 $\pm$ 0.66 & 0.79 $\pm$ 1.22 & 3.38 $\pm$ 4.25 & 28.5\\
 & 0.73 $\pm$ 0.49 & 0.86 $\pm$ 0.41 & 2.33 $\pm$ 0.25 &  \\
\hline
Radio+p1 & 1.55 $\pm$ 0.66 & 1.11 $\pm$ 0.44 & 2.80 $\pm$ 0.46 & 31.3\\
 & 0.50 $\pm$ 0.40 & 0.72 $\pm$ 0.29 & 2.21 $\pm$ 0.36 &  \\
\hline
X-ray+p1 & 1.31 $\pm$ 0.60 & 0.88 $\pm$ 0.38 & 2.81 $\pm$ 0.36 & 30.3\\
 & 0.59 $\pm$ 0.41 & 0.77 $\pm$ 0.29 & 2.27 $\pm$ 0.33 &  \\
\hline
$\rm p_{center}+p1$ & 1.11 $\pm$ 0.56 & 0.58 $\pm$ 0.64 & 3.57 $\pm$ 4.67 & 28.9\\
 & 0.79 $\pm$ 0.44 & 0.89$\pm$ 0.35 & 2.36 $\pm$ 0.28 &  \\
\hline
\hline
\end{tabular}
\begin{tablenotes}
\item\textbf{Notes.} The Disk model is a uniform disk with a radius corresponding to the virial radius  $\theta_{200}$. The Core model is a predicted gamma-ray flux profile (see the text for details). The point source model A and B correspond to point sources at the center of the Coma cluster and at the position of p1 in \figurename~\ref{fig1}, respectively. In each two-component model,  the flux and photon spectral index of the each component are listed in the top/bottom line. The associated uncertainty refers to the 68\% error reported by HESSE algorithm embedded in the {\it gtlike} tool.
\end{tablenotes}
\label{taba1}
\end{threeparttable}
\end{table*}

\section {Estimate of the CR to thermal energy ratio}
The collision rate between CR protons with the ICM is
\begin{equation}
\frac{dN_c}{dt}=\sigma_{pp} n_g c n_{\rm cr}
\end{equation}
where $\sigma_{pp}\simeq 2.5\times10^{-26} {\rm\ cm^2}$ is the interaction cross section for $pp$ collisions at GeV energies, $n_g$ is the number density of the ICM and $n_{\rm cr}$ is the number density of CR protons at energy $\epsilon_{\rm cr}$. Roughly, the collisions produce two photons of energy
$\epsilon_\gamma=\frac{1}{2}{\kappa}\epsilon_{\rm cr}$, where $\kappa\sim 0.17$ is the fraction energy transferred from the proton to secondary pions.
The total gamma-ray luminosity  within the virial radius $R_{vir}$ is \cite{2013A&A...560A..64H}
\begin{equation}
\epsilon_\gamma L_{tot}(\epsilon_\gamma)=\int_0^{R_{vir}}2\epsilon_\gamma \sigma_{pp}n_g(r) n_{\rm cr}(r) c 4\pi r^2 dr.
\end{equation}
Assuming CR number follows a power-law distribution with the form $n_{\rm cr}(\epsilon_{\rm cr})=n_{\rm cr}(\epsilon_0)(\frac{\epsilon_{\rm cr}}{\epsilon_0})^{-p+1}$ (${\epsilon_{\rm cr}}\ge{\epsilon_0}=1{\rm GeV}$), the energy density of CRs is $U_{\rm cr}=\int n_{\rm cr} d\epsilon_{\rm cr}$. The normalization of the CR energy density can be obtained by assuming a constant CR to thermal energy density ratio, $U_{\rm cr}=f_{\rm cr}U_{\rm th}$. The energy density of thermal gas of $U_{\rm th}=\frac{3}{2} n_g k T$, where $T=10^8 {\rm\ K}$ is the temperature of the ICM of the Coma cluster. For simplicity, we assume that that the density of the thermal gas follows an isotheral beta model, which is $n_g=n_0 [1+(r/r_c)^2]^{-3\beta/2}$, where $n_0$ is the central density, $r_c$ is the cluster core radius and $\beta$ is the slope of the density profile outside of the core. We take $n_0=3\times10^{-3}{\rm\ cm^{-3}}$, $r_c=290{\rm\ kpc}$  and $\beta=2/3$ for the Coma cluster.
According to Table I, the differential luminosity of the Coma cluster at 200 MeV is about $\epsilon_\gamma L_{tot}(200{\rm MeV})\simeq10^{42}{\rm\ erg\ s^{-1}}$. Comparing this with the above equation, we obtain that the volume averaged value of the CR to thermal energy ratio is
\begin{equation}
f_{\rm cr}=\frac{U_{\rm cr}}{U_{\rm th}}\simeq 2\% .
\end{equation}


\end{document}